\documentclass[12pt]{iopart}
\usepackage{iopams}  
\usepackage{epsfig}
\begin{document}
\title[System size dependence of strangeness production]
{System size dependence of strangeness production at 158 $A$\hspace{1mm}GeV}
\author{I.~Kraus for the NA49 Collaboration
\footnote[3]{see full authors list in the contribution of V. Friese in this volume}}
\address{Gesellschaft f\"{u}r Schwerionenforschung (GSI), Darmstadt, Germany.}
\begin{abstract}
An enhanced production of strange particles per pion in heavy ion collisions 
compared to elementary p+p interactions was observed in Pb+Pb reactions at 
the CERN SPS. New results obtained on small colliding systems, C+C and Si+Si, 
also show an enhancement. It increases steeply with system size
and reaches in the larger system almost 
the level measured in Pb+Pb reactions.
The shapes of the transverse mass spectra of strange
mesons (K$^+$, K$^-$, $\phi$) and baryons ($\Lambda$, $\bar{\Lambda}$)
are consistent with the presence of collective transverse
flow, which is already visible in small systems (C+C) and increases with the system size.
The $\bar{\Lambda}/\Lambda$ ratio together with the $K^{+}/K^{-}$ ratio
yields information about the baryochemical
potential $\mu_{B}$. The potential, determined at midrapidity, increases with the 
system size since the stronger stopping shifts baryons from near beam and 
projectile rapidity towards midrapidity. For the $4 \pi$ yields $\mu_{B}$ is almost 
independent of the size of the colliding system.
\end{abstract}
\vspace{-7mm}
\section{Introduction}
Strangeness enhancement in heavy ion reactions compared to elementary interactions
and strong radial expansion of the collision fireball were treated, among others, as indications 
for the formation of deconfined strongly interacting matter \cite{Heinz}. Both were observed in Pb+Pb
reactions at the CERN SPS \cite{pbpb}. Prior results from sulfur induced reactions at 
200 $A$\hspace{0.6mm}GeV \cite{na35} also showed a strangeness enhancement.
These results provoked questions about the evolution of strangeness production and radial flow with the
size of the fireball.

Central collisions of small systems will be compared to previously measured p+p and 
central Pb+Pb interactions instead of centrality selected Pb+Pb reactions, because it was
recently shown in the framework of Glauber and UrQMD calculations \cite{hoe} 
that the peripheral collision geometry leads to a different fraction of nucleons suffering
multiple collisions and to different space time densities of inelastic 
collisions compared to central reactions of small nuclei at the same number of wounded nucleons.
\section{Experimental procedure}
\begin{figure}
\begin{center}
\includegraphics[scale=0.63]{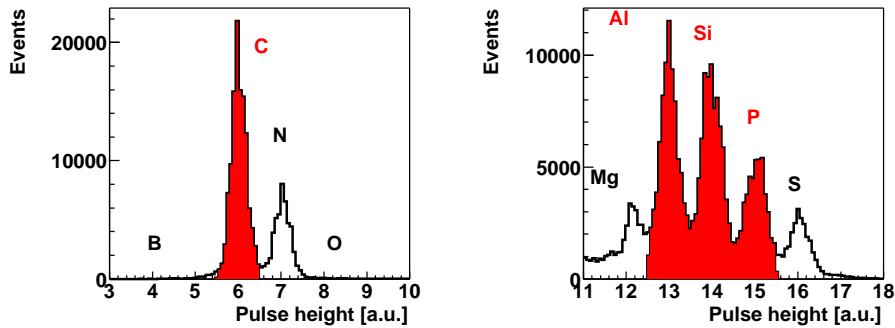}
\vspace{-5mm}
\caption{\label{label1}Pulse height distributions of the Pb fragmentation beam, measured in beam line
detectors. The dark areas indicate the selected beam particles.}
\end{center}
\end{figure}
In this section the definition of the C and Si beams and the event statistics
as well as the experimental setup and the particle identification procedures are described.

The CERN SPS accelerator provided a 158 $A$\hspace{0.6mm}GeV Pb beam, which was steered onto 
a C fragmentation target. The fragments were selected in the beam line according to their rigidity. 
Figure \ref{label1} shows the pulse height distributions, which correspond to the charge distributions of 
the beam particles, selected online by additional energy loss cuts,
for the C setting (left) and Si (right).
The final identification
was carried out offline by charge and energy loss measurements obtained from beam line detectors.
The pion, kaon and $\phi$ meson analysis is based on
a mixed beam of C and N ions, while the pure C beam was used for the hyperon analysis.
Both will be called C+C later on, the Si+Si named results arise from a mixed Al, Si and P beam
for all analysed particles.

The data sets consist of 46k C+C events, taken with a 2.4$\%
$ interaction length C target (0.56 $\rm g/cm^2$) for the meson analysis. In addition a $7.9\%
$ interaction length C target (1.84 $\rm g/cm^2$) was used to increase the statistics to 252k events for
the hyperon analysis. The Si target (1.17 $\rm g/cm^2$) had an interaction length of $4.4\%
$, 43k events were used for the mesons analysis and 173k events in case of the hyperons.
The centrality selection was performed by an online trigger on the energy deposited in the 
zero degree calorimeter, which measures the energy of the projectile spectators.
The $ 17.5 \pm 1.5 \%
 (12.5 \pm 1.5 \%
)$ most central C+C (Si+Si) events were selected. The corresponding 
numbers of wounded nucleons 
$(<N^{C+C}_{W}> = 16 \pm 1, <N^{Si+Si}_{W}> = 41.5 \pm 1.5)$
are derived from VENUS simulations \cite{venus}. 

The NA49 large acceptance hadron spectrometer \cite{nim} consists of four time projection chambers,
two of them are placed inside a magnetic field downstream of the target
and two further downstream on either side of the beam line.
Tracking of charged particles 
in the forward hemisphere ($y > y_{cm} = 2.9$) and 
measurement of the specific energy loss (dE/dx) give access to the momentum and particle 
identification.
The $\phi$ mesons were identified by calculating the invariant mass of $K^+ K^-$ pairs,
for the hyperons ($\Lambda \rightarrow p \pi^-, \bar{\Lambda} \rightarrow \bar{p} \pi^+$) 
requirements on the invariant mass and decay topology were used.
Losses due to geometrical acceptance and reconstruction inefficiency were determined from GEANT 
simulations. The simulations were also used to correct for 
$K^+$ and $K^-$ decay in flight and $\Lambda$ 
and $\bar{\Lambda}$ finding inefficiency and losses due to restrictions in the decay topology. 
Corrections for the feeding from weak decays are not yet applied.
\section{Particle yields and spectra}
\begin{figure}
\begin{center}
\includegraphics[scale=0.60]{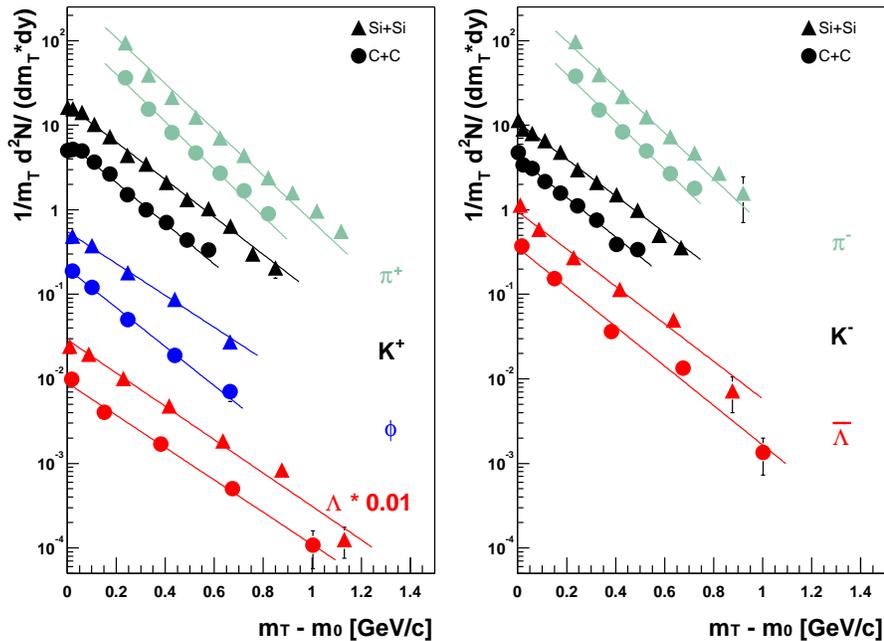}
\vspace{-5mm}
\caption{\label{label3}Transverse mass spectra at midrapidity. Triangles are used for particles 
produced in Si+Si interactions, circles for C+C reactions at 158 $A$\hspace{0.6mm}GeV.
The spectra are integrated over  the 
rapidity interval $3.0 < y < 3.23$ for $\pi$ and K, 
$2.5 < y < 2.9$ for $\Lambda$ particles.
Due to low statistics, a wider rapidity interval had to be used for $\phi$ mesons ($2.9 < y < 4.7$).
The lines represent the exponential fits.}
\end{center}
\end{figure}
\begin{figure}
\begin{center}
\includegraphics[scale=0.53]{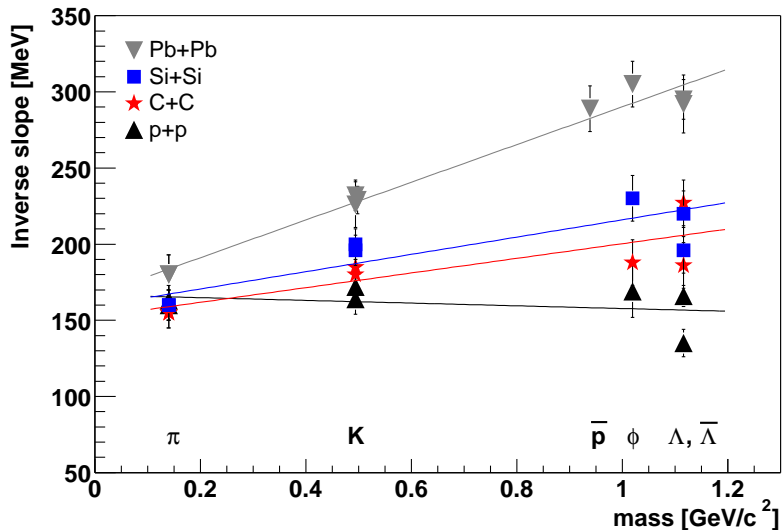}
\vspace{-5mm}
\caption{\label{label4}Inverse slope parameter as a function of the particle mass for different collision
systems at 158 $A$\hspace{0.6mm}GeV.}
\end{center}
\vspace{-5mm}
\end{figure}

Particle yields of charged pions and kaons, $\phi$ mesons and hyperons
in C+C and Si+Si collisions were obtained in rapidity and $p_T$ bins.
The transverse mass spectra ($m_T = \sqrt{p_T^2 + m_0^2}$) measured close to midrapidity 
are shown in figure \ref{label3}. Motivated by the thermal ansatz
they were fitted with an exponential 
\begin{equation*}
\frac{d^2N}{m_T \cdot dy \cdot dm_T} \sim e^{-m_T/T}.
\end{equation*}
The fitted inverse slope parameters $T$
are shown in figure \ref{label4} as a function of the particle mass. Results of p+p \cite{pp} and
Pb+Pb reactions \cite{pbpb} are included 
for comparison. The lines are meant to guide the eye, their increasing slope can be 
interpreted as increasing radial flow, which is not visible in p+p interactions, but
already present in C+C interactions and further rising with the system size.
This is an indication of collective effects already established in small systems.
The inverse slope parameter of $\Lambda$ hyperons is 20 $\%
$ above the one of $\bar{\Lambda}$ hyperons in p+p interactions, this difference is decreasing 
with the system size and vanishes in the largest colliding system.

The exponential fits were used to extrapolate the measured transverse mass spectra in order to obtain
rapidity densities, which were finally fitted by single Gaussians for the pions, kaons and $\phi$ mesons. 
The integration results in the total yields of the various particles (see \cite{hoe} for details).
The rapidity distributions for $\bar{\Lambda}$ hyperons, shown in
figure \ref{label7}a, were fitted by a single Gaussian as well, but because of the lack of data far-off
midrapidity there is a large uncertainty in the extrapolation. This leads to large systematic
errors. The total multiplicities amount to 
$<\bar{\Lambda}>_{C+C} = 0.27 \pm 0.06(stat) \pm 0.10(sys)$ and
$<\bar{\Lambda}>_{Si+Si} = 0.66 \pm 0.20(stat) \pm 0.33(sys)$, respectively.
The extrapolation of the
rapidity distributions of $\Lambda$ hyperons, shown in figure \ref{label7}b,
is even more complicated, because the almost flat distributions provide no indication for the shape in 
the not measured rapidity region. Two extreme cases were assumed to estimate upper and lower
limits of the yields. The curve on the right hand side (figure \ref{label7}b)
is a scaled parametrisation of the $\Lambda$
rapidity distributions measured in p+p interactions. The forward/backward peaked shape provides
an upper limit of the yield. The straight line on the left hand side, obtained from
the scaled distribution measured in S+S reactions, should create a lower limit, since the hyperons are expected to be 
shifted towards midrapidity due to stronger stopping in S+S compared to the smaller C+C and Si+Si systems. 
The yields, calculated as the average of both extrapolation methods are
$<\Lambda>_{C+C} = 1.43 \pm 0.05(stat) \pm 0.25(sys)$ and
$<\Lambda>_{Si+Si} = 4.50 \pm 0.16(stat) \pm 0.84(sys)$, respectively.

\section{System size dependence}
\begin{figure}
\begin{center}
\includegraphics[scale=0.69]{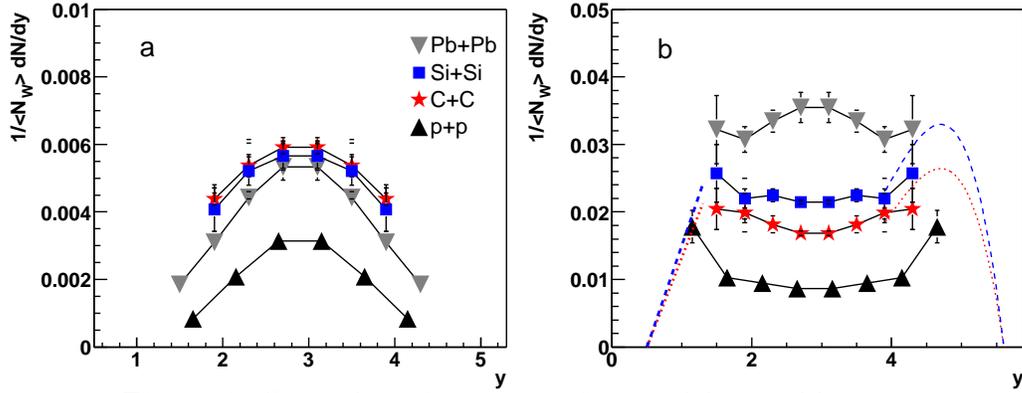}
\vspace{-5mm}
\caption{\label{label7}Comparison of the shape of the rapidity densities per wounded nucleon,
measured in different collision systems. 
The left figure shows $\bar{\Lambda}$ and the right one $\Lambda$ hyperons.
The rapidity distributions are forward/backward averaged, 
the lines in the right figure show the different assumptions for the extrapolation to total yields
in C+C and Si+Si reactions.}
\end{center}
\end{figure}
\begin{figure}
\vspace{-5mm}
\begin{center}
\includegraphics[scale=0.55]{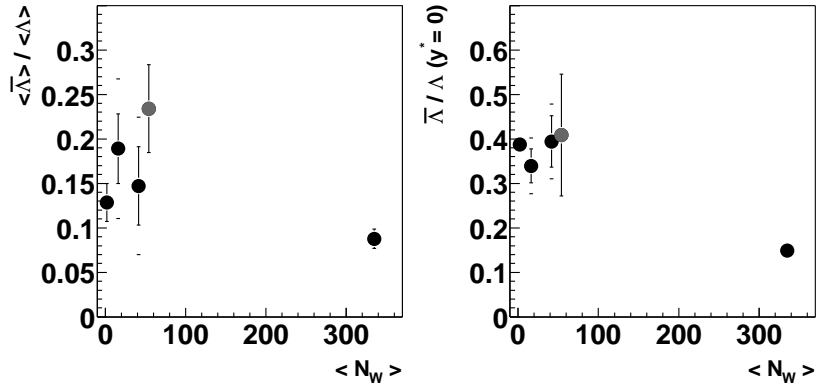}
\vspace{-5mm}
\caption{\label{label10}$\bar{\Lambda}/\Lambda$ ratio of the total yields (left) and
at midrapidity (right) obtained in p+p, C+C, Si+Si and Pb+Pb collisions at 158 $A$\hspace{0.6mm}GeV 
and S+S reactions at 200 $A$\hspace{0.6mm}GeV.
}
\end{center}
\vspace{-5mm}
\end{figure}
The increasing baryon stopping with the system size, mentioned above, is illustrated in figure \ref{label7},
where the rapidity spectra of $\bar{\Lambda}$ and $\Lambda$ hyperons, measured in several
collision systems are compared.
The newly produced $\bar{\Lambda}$ hyperons are always peaked at midrapidity 
and their width is not affected by the increasing probability of multiple collisions, while,
especially in the small systems, an essential fraction of the $\Lambda$ hyperons carry quarks
(momentum) of the beam and projectile nucleons. So, their evolution from the forward/backward 
peaked shape to a maximum at midrapidity with the size of the fireball is a clear indication for
increasing stopping of the incident nucleons.

This has a strong effect on the $\bar{\Lambda} / \Lambda$ ratio at midrapidity (figure \ref{label10} right),
which is decreasing by more than a factor of two  when going from p+p interactions to Pb+Pb collisions. 
The ratio of the total yields (figure \ref{label10} left) exhibit a less pronounced dependence on the system 
size. A similar behavior is seen for the $K^- / K^+$ ratios (not shown), both at midrapidity and in full phase
space.
The higher value of the ratio in S+S reactions might be due to the lower 
stopping at higher beam energy.
The clear difference between the system size dependence of midrapidity and $4\pi$ ratios has a significant 
impact on the comparison of the data with statistical models.
In this framework the $\bar{\Lambda} / \Lambda$ ratio together with the $K^{+}/K^{-}$ ratio is a measure of 
the baryochemical potential, since they are related by 
$(\bar{\Lambda} / \Lambda)^{-1/6} \cdot (K^{+}/K^{-})^{1/6} \hspace{7mm}\lsimeq\hspace{0.5mm} 
exp(1/3 \cdot \hspace{0.5mm} \mu_B/T_{chem})$ 
\cite{model}
and the chemical {\it freeze-out} temperature can be assumed, as a good approximation, to be about 
160 MeV for all systems \cite{bec}. 
The potential, calculated from the total yields, is in the range of
190 MeV and shows only a weak dependence on the system size.
This is slightly below statistical model fits \cite{cley} based on different particle yields except the 
hyperons, while $\mu_B$, deduced from midrapidity $\bar{\Lambda} / \Lambda$ ratios, reaches just 110 MeV 
in the small systems.
\begin{figure}
\begin{center}
\includegraphics[scale=0.78]{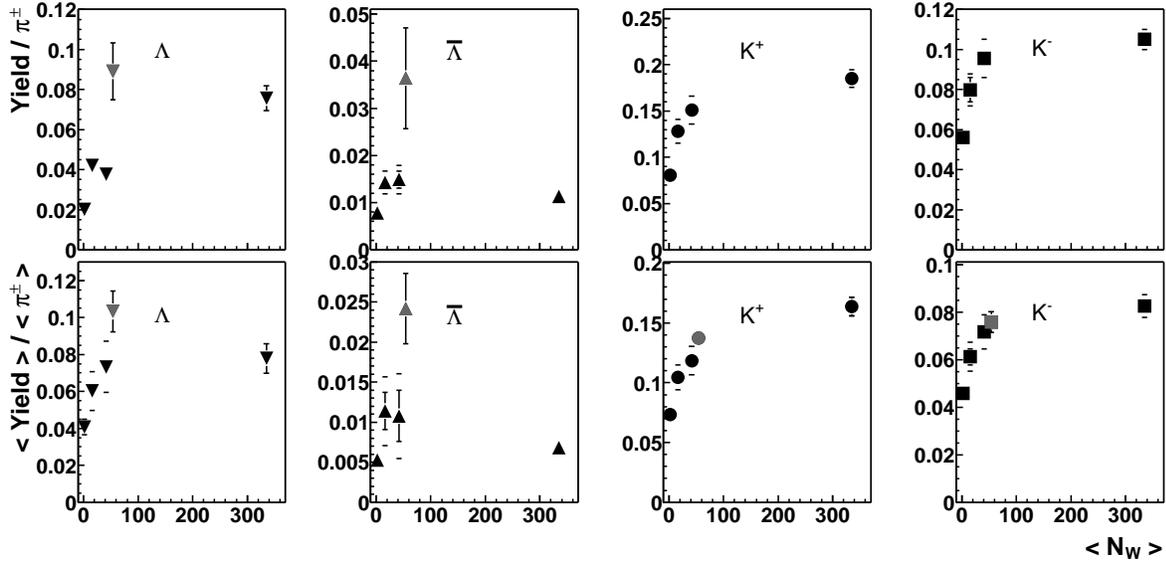}
\vspace{-5mm}
\caption{\label{label8}Midrapidity yields per $\pi$ vs. the number of wounded nucleons
in the upper row and total yields per $\pi$ vs. the number of wounded nucleons in the lower row.
Obtained in p+p, C+C, Si+Si, S+S and Pb+Pb collisions.}
\end{center}
\end{figure}

The yield of strange particles per pion ($\pi^\pm = 1/2 (\pi^+ + \pi^-)$) as a function
of the number of wounded nucleons is shown in figure \ref{label8} for the midrapidity densities (top)
and for the total multiplicities (bottom).
The relative K$^+$, K$^-$ and $\Lambda$ production is steeply rising with the system size,
most of the enhancement obtained in Pb+Pb collisions is already reached in Si+Si and S+S interactions.
This behaviour is expected in the thermal picture for
the transition from canonical to grand canonical ensembles. 
In the same picture the maximum in the $\bar{\Lambda} / \pi$ ratio again at intermediate system sizes
can be attributed to the increasing $\bar{\Lambda}$ production
because of this volume effect
followed by a decrease due to the larger baryochemical potential
in larger systems.
\end{document}